\documentstyle[12pt]{article}

\def\GeV{\,{\rm GeV}}

\def\keV{\,{\rm keV}}
\def\MeV{\,{\rm MeV}}
\def\sec{\,{\rm sec}}
\def\Gyr{\,{\rm Gyr}}

\def\Mpc{\,{\rm Mpc}}

\def\eV{{\,\rm eV}}

\def\cmm2{{\,\rm cm^{-2}}}
\def\cm2{{\,{\rm cm}^2}}
\def\cmm3{{\,{\rm cm}^{-3}}}
\def\gcmm3{{\,{\rm g\,cm^{-3}}}}
\def\kms{\,{\rm km\,s^{-1}}}

\def\mpl{{m_{\rm Pl}}}

\def\he{$^4$He}

\def\la{\mathrel{\mathpalette\fun <}}

\def\fun#1#2{\lower3.6pt\vbox{\baselineskip0pt\lineskip.9pt
  \ialign{$\mathsurround=0pt#1\hfil##\hfil$\crcr#2\crcr\sim\crcr}}}

\begin{document}
\pagestyle{empty}
\begin{center}
\bigskip
\rightline{FERMILAB--Pub--94/026-A}
\rightline{astro-ph/9402028}
\rightline{submitted to {\it Physical Review Letters}}

\vspace{.2in}
{\Large \bf IS A MASSIVE TAU NEUTRINO\\
\medskip
JUST WHAT\\
\bigskip
COLD DARK MATTER NEEDS?} \\

\vspace{.2in}
Scott Dodelson,$^1$ Geza Gyuk,$^{1,2}$ and Michael S. Turner$^{1,2,3}$\\
\vspace{.2in}

{\it $^1$NASA/Fermilab Astrophysics Center\\
Fermi National Accelerator Laboratory, Batavia, IL~~60510-0500}\\

\vspace{.1in}

{\it $^2$Department of Physics\\
The University of Chicago, Chicago, IL~~60637-1433}\\

\vspace{0.1in}
{\it $^3$Department of Astronomy \& Astrophysics\\
Enrico Fermi Institute, The University of Chicago,
Chicago, IL~~60637-1433}\\

\end{center}

\vspace{.3in}

\centerline{\bf ABSTRACT}
\medskip

\noindent  The cold dark matter (CDM) scenario for structure
formation
in the Universe is very attractive and has many successes;
however, when its spectrum of density perturbations is
normalized to the COBE anisotropy
measurement the level of inhomogeneity predicted on small
scales is too large.
This can be remedied by a tau neutrino of mass $1\MeV - 10\MeV$
and lifetime $0.1\sec - 100\sec$ whose decay products include
electron neutrinos because it allows the total energy density
in relativistic particles to be doubled without interfering
with nucleosynthesis.  The anisotropies predicted
on the degree scale for ``$\tau$CDM'' are
larger than standard CDM.  Experiments at $e^\pm$ colliders may
be able to probe such a mass range.

\newpage
\pagestyle{plain}
\setcounter{page}{1}

The cold dark matter (CDM) scenario for the formation of structure
in the Universe (i.e., galaxies, clusters of galaxies,
superclusters,
and so on) is both well motivated and very successful
\cite{ucsc}.  The CDM model begins with a flat Universe with
most of its mass in slowly moving (``cold'') particles
such as axions or neutralinos and a baryon fraction compatible
with primordial nucleosynthesis ($\sim 5\%$).  The density
perturbations that seed structure formation
are the nearly scale-invariant perturbations that arise from
quantum fluctuations during inflation \cite{scalar}.  Standard
CDM has one free parameter:  the normalization of the spectrum.
The recent COBE detection \cite{DMR} of anisotropy in the
temperature of the
cosmic background radiation (CBR) now provides the normalization
by fixing the amplitude of perturbations
on very large scales ($\lambda\sim 10^3h^{-1}\Mpc$).
But, when the spectrum is so normalized, the level
of inhomogeneity predicted on small scales
($\lambda \la 10h^{-1}\Mpc$) is too large by a factor of two
(the Hubble constant $H_0 =100h \kms \Mpc^{-1}$).
Just how serious this problem is remains to be seen.

A number of ``fixes'' have been proposed.  They involve
changing either the initial power spectrum or the energy content of
the Universe.  The simplest way of changing the power spectrum
is ``tilt'' \cite{tilt} (i.e., deviation from scale invariance)
which
reduces the power on small scales; in fact, several plausible
models of
inflation predict a tilted spectrum.  However,
this solution seems to lead to insufficient power on
intermediate scales ($\lambda \sim 30h^{-1}
\Mpc - 100h^{-1}\Mpc$).  Changing
the energy content changes the transfer function, which relates
the primeval power spectrum to that today \cite{transfer}.
For example, mixed dark matter (MDM) \cite{mdm} (65\% CDM + 30\%
hot dark matter in the form of $7\eV$ neutrinos + 5\% baryons) or
$\Lambda$CDM
\cite{lcdm} (80\% vacuum energy + 15\% CDM + 5\% baryons) both
reduce the power on small scales.

The fix advocated here involves increasing the energy
density in relativistic particles, which also modifies the
transfer function.
In the standard scenario the radiation content at late times
($t\gg 1\sec$) consists of photons and three massless neutrino
species with slightly lower temperature
$T_\nu = (4/11)^{1/3}T_\gamma$,
accounting for a total radiation energy density $\rho_{\rm rad} =
g_*{\pi^2}{T_\gamma}^4/30$.  Here
$g_* =2 + 2(7/8)(4/11)^{4/3}N_\nu$
counts the number of effectively massless degrees of freedom and
is equal to
$3.36$ for $N_\nu =3$.  In our ``$\tau$CDM'' model $g_*$ will
be about
a factor of two larger.

To see why this helps, consider the scale $\lambda_{\rm EQ}$, which
crossed the horizon at the time the energy density
in matter was equal to that in radiation.
On scales much greater than $\lambda_{\rm EQ}$ the transfer
function is unity because perturbations on
these scales crossed the horizon during the
matter-dominated era and began growing as soon as they did.
 On scales
much smaller than $\lambda_{\rm EQ}$ the transfer function
decreases as $\lambda^2$ because perturbations on these
scales crossed the horizon before matter-radiation
equality and did not begin growing until the matter-dominated era.
Thus, $\lambda_{\rm EQ}$ controls the
shape of the transfer function; it is given by,
$\lambda_{\rm EQ} \sim (R_0/R_{\rm EQ})t_{\rm EQ}$
($R_0$ is the scale factor today; $R_{\rm EQ}$ and $t_{\rm EQ}$
are the scale factor and age of the Universe at matter-radiation
equality).  Since the energy density in matter red shifts as
$R^{-3}$ and
that in radiation as $R^{-4}$,
$R_0/R_{\rm EQ} = \rho_{\rm matter}/
\rho_{\rm rad} \simeq 2.4\Omega_0 h^2\times 10^4/ (g_*/3.36)$
($\Omega_0$ is the ratio of the energy density
to the critical density; $\Omega_0 =1$ for a flat Universe).
The age
at matter-radiation equality $t_{\rm EQ} \sim
\mpl /g_*^{1/2}T_{\rm EQ}^2$, and so
\begin{equation}
\lambda_{\rm EQ} \sim 10\Mpc (\Omega_0h^2)^{-1}(g_*/3.36)^{1/2}.
\end{equation}
Increasing $g_*$ increases $\lambda_{\rm EQ}$; since
COBE fixes the power spectrum on scales
$\lambda \gg \lambda_{\rm EQ}$, doing so reduces
the power on small scales (see Fig.~1).

How much additional radiation is needed?
Since cosmological distances scale as the inverse
of the Hubble constant, the shape of the power spectrum
is set by $(h^{-1}\Mpc )/ \lambda_{\rm EQ} \propto \Gamma \equiv
(\Omega_0 h)(g_*/3.36)^{-1/2}$.  The best
fit to all the data seems to require a ``shape parameter''
$\Gamma \simeq 0.3$ \cite{shape}, whereas $\Gamma \simeq 0.5$ for
the canonical values of $\Omega_0 = 1.0$, $g_* = 3.36$, and
$h=0.5$.\footnote{In order
to insure a sufficiently aged Universe with $\Omega_0 =1$,
$h$ must lay at the lower end of current
measurements, $h\sim 0.5$.  Of course, another
fix is to posit a very low value of the Hubble parameter,
$h=0.3$, which leads to a very old Universe, $t_0 \simeq 22\Gyr$.}
For a given Hubble constant and shape parameter,
the number of equivalent neutrino species called for is
$N_\nu =  20 (0.3/\Gamma )^2\Omega_0^2(h/0.5)^2 - 4.3$.
To achieve $\Gamma = 0.3$ with $h=0.5 (0.4)$ requires
$N_\nu = 16 (8.5)$.

Therein lies the rub:  Primordial nucleosynthesis restricts the
equivalent number of massless neutrino species to be less than 3.4
($g_* \le 3.54$) \cite{walker}.  The argument is well known:
More relativistic degrees of freedom increase the expansion
rate during nucleosynthesis and lead to an earlier
freeze out of the neutron fraction at a higher value;
this results in too much $^4$He if $N_\nu > 3.4$.

One way around this has been discussed previously:
a late-decaying, massive particle species, e.g., the once
viable $17\keV$ neutrino \cite{sjf}.  Since the energy density
in a massive particle species grows relative to radiation
as the scale factor, a massive particle
whose energy density at the epoch of nucleosynthesis is negligible
can produce a significant amount of radiation
if it decays late enough, in the case of the $17\keV$ neutrino,
$t\sim 10^7\sec$ \cite{be}.

Our proposal is very different and involves a subtle
aspect of nucleosynthesis.  We recently examined the effect
of an unstable tau neutrino on nucleosynthesis for
decay modes including $\nu_\tau \rightarrow \nu_e +\phi$
($\phi$ is
an effectively noninteracting scalar particle) \cite{dgt}.
The decay products of a massive tau neutrino can have
more energy density than a massless neutrino species; for
most decay modes
this leads to an unacceptable increase in \he.  However, for
the $\nu_e\phi$ decay mode there is a competing effect:
Some decay-produced $\nu_e$'s are captured by neutrons,
reducing the neutron fraction and suppressing \he\
production \cite{satoeffect}.
This opens the door for adding relativistic degrees of
freedom---which we've seen can help CDM---without
overproducing $^4$He.  The effect
is even more pronounced for the $\nu_\tau
\rightarrow \nu_e+\nu_e{\bar\nu}_e$ mode.

The key to all of this is the fact that the decay-produced
$\nu_e$'s resemble a bath of neutrinos with
temperature lower than that at which the neutron
fraction freezes out ($\sim 1\MeV$), and thus, their interactions
with nucleons tend to reduce the neutron fraction.
The rate for neutron-proton interconversion
($n+\nu_e \leftrightarrow p+e^-$, etc.) varies as $T^5$ and so
the typical energy of a captured neutrino is $5T$, or $5\MeV$
at freeze out; decay-produced electron neutrinos have energy
 $m_\nu /2$
(lower for the $3\nu_e$ mode).  Thus,
the effective temperature of decay-produced neutrinos
will be lower than $1\MeV$ provided $m_\nu \la 10\MeV$.
Conversely, for masses greater than $10\MeV$, decay-produced
neutrinos
increase the neutron fraction as they appear hotter.
This simple explanation is borne out by detailed calculations
\cite{dgt}.

Next, consider the lifetime dependence.  Decays much earlier than
$1\sec$ occur before freeze out, when the neutron fraction
is close to its equilibrium value and so there is little effect.
The dilution effect of the expansion decreases the
capture probability of a $\nu_e$, and so few
$\nu_e$'s produced much later than $1\sec$ are
captured.  Furthermore, the energy density of the massive tau
neutrino grows relative to the radiation as the scale factor,
and for lifetimes greater than $10\sec$ increases the expansion
significantly rate, tending to increase $^4$He production.
For long lifetimes the increase in expansion rate wins out
and leads to higher $^4$He production; for lifetimes around a
few seconds the two competing effects cancel; for short lifetimes
the effect of neutron captures wins out leading to lower
$^4$He production.   For a lifetime of $\sim 0.3\sec$
the reduction
in $^4$He is maximized and can be as large
$\Delta Y_P \sim 0.10$ \cite{dgt}.

There are three ways to take advantage of the ``nucleosynthesis
physics'' described above to increase
the radiation content in the Universe.  The first involves
a tau neutrino of mass $1\MeV -10\MeV$ and lifetime $0.1\sec
-{\rm few}\sec$.  Tau-neutrino decays can reduce the neutron
fraction enough to make room for up to the equivalent of 16
additional
neutrino species (see Fig.~2).  For lifetimes this short
the energy density of the tau neutrino does not become
significant before it decays, and the extra energy density in
relativistic
species must arise from additional massless degrees of freedom.
Many extensions of the standard model predict
new massless degrees of freedom (see below).
(While we have not explored masses less than $1 \MeV$ for technical
reasons \cite{dgt}, Fig.~2 suggests that a massive muon neutrino
that decays to electron neutrinos might also allow for many
additional
massless species.)

The second possibility involves a tau neutrino of lifetime
$10\sec -
50\sec$ and is probably the most attractive.
Here, not only do tau-neutrino decays inhibit $^4$He production
but they also produce the additional energy density in
relativistic
particles.  That energy density, expressed in equivalent number
of neutrino species, is roughly \cite{kt}
\begin{equation}\label{eq:delta}
\Delta N_\nu \simeq 0.4 (rm_\nu /\MeV)(\tau_\nu /\sec )^{1/2},
\end{equation}
where $r$ is the freeze-out abundance of
the massive tau neutrino relative to that of a
massless neutrino species ($r\sim 0.1 -1$ for $m_\nu =
1\MeV - 10\MeV$) and the total equivalent number of
massless neutrinos $N_\nu = 2+\Delta N_\nu$.

In Fig.~3 we show the regions of the tau-neutrino mass-lifetime
plane that are excluded by primordial nucleosynthesis and
contours of the energy density of the tau-neutrino's decay
products (based on our numerical calculations \cite{dgt}).
We exclude a mass/lifetime if for no value of the baryon-to-photon
ratio the light-element abundances satisfy:  D/H $\ge 10^{-5}$;
(D+$^3$He)/H $\le 1\times 10^{-4}$;
$Y_P \le 0.24$; $^7$Li/H $\le 1.4\times 10^{-10}$.
As discussed earlier, this excludes long lifetimes.
Since $\Delta N_\nu$ grows with lifetime, the boundary of the
forbidden region is most interesting.  Since it
depends most sensitively on the limit to $Y_P$ (and to a lesser
degree on $^7$Li) we also show the boundary for the
relaxed constraints, $Y_P\le 0.25$ and $^7$Li/H $\le 2\times
10^{-10}$.  For a Dirac neutrino
$\Delta N_\nu \simeq 5 (7)$ ($\nu_e\phi$) and
$8(10)$ ($3\nu_e$) are the maximum permitted
(the number in parenthesis for the relaxed constraints).
The $\Delta N_\nu$'s for a Majorana neutrino are about 2
neutrino species smaller as their abundance is smaller \cite{dgt}.

The final possibility is a variation on the previous one.
While a massive tau neutrino provides a neat way
of simultaneously suppressing $^4$He production and
producing additional radiation, a species with similar
decay modes and a higher abundance could
do even better.  As noted earlier, for a lifetime of a few seconds
the effects of decay-produced $\nu_e$'s and increased
expansion rate cancel and increasing
$r$ does not affect $^4$He production---but does increase
$\Delta N_\nu$, cf. Eq.~(\ref{eq:delta}).  Thus for
$\tau \sim 0.1\sec
- {\rm few}\, \sec$ $\Delta N_\nu$ can be increased
almost without limit.  While $r$ is not too different from
one for a tau neutrino of mass $1\MeV -10\MeV$,
it could be larger for a species with more
degrees of freedom or weaker interactions.  To illustrate,
for a hypothetical $5\MeV$ species with $r=2$,
$\Delta N_\nu = 10$ ($\nu_e\phi$), 17 ($3\nu_e$) as
large as is permitted.

Next, we consider a few of the consequences of $\tau$CDM.
The level of inhomogeneity on small scales can
be quantified by {\it rms} mass fluctuation in spheres
of radius $8h^{-1}\Mpc$, $\sigma_8$.  For COBE-normalized, standard
CDM $\sigma_8 \simeq 1.2$; for mixed dark matter
$\sigma_8 \simeq 0.75 $; and for $\tau$CDM model with
$\Gamma = 0.3$, $\sigma_8 \simeq 0.67 $.  The level of
inhomogeneity predicted in $\tau$CDM on small scales is about a
factor of two smaller than CDM and slightly lower than MDM.
Since the {\it rms} fluctuation in galaxy counts is unity
on the $8h^{-1}\Mpc$ scale, light
is a ``biased'' tracer of mass in both $\tau$CDM
and MDM, with bias factor $b\equiv 1/\sigma_8 \simeq 1.5$.
On the other hand, as can be seen in Fig.~1, $\tau$CDM has
more power than MDM on very small scales and so forming
objects such as quasars at red shifts $z\sim 4-5$ should be
less problematic.

We have computed the angular power spectrum of CBR anisotropy,
$C_l = \langle |a_{lm}|^2\rangle$ ($\delta T =\sum a_{lm}
Y_{lm}$), and it is larger than standard CDM (see Fig.~4).
To understand why, recall
that the extra degrees of freedom delay matter-radiation
equality.  Last scattering occurs at the
time of recombination ($T\sim 1/3\eV$), and in $\tau$CDM
this happens closer to---or perhaps even in---the
 radiation-dominated era.
As a result, the gravitational potential photons move
through as they travel from the last-scattering surface
is not constant, which leads to larger CBR anisotropy \cite{scott}.
The quantity $l(l+1)C_l$ roughly corresponds to the expected
variance of the CBR temperature difference measured between
directions separated by angle $\theta \sim (200/l)^\circ$.
For $l\sim 100-300$ this quantity is about a factor of
1.6 larger than standard CDM, which implies predicted
temperature anisotropies on the degree scale that are a factor
of $\sqrt{1.6}$ larger.  While the situation is far from
settled, there are experiments with detections that are
larger than the CDM prediction \cite{degree}.

In passing we mention two more consequences.  First,
a tau neutrino of mass $1\MeV -10\MeV$
can resurrect a theory that predicts too many massless
degrees of freedom to be consistent with the ``standard
nucleosynthesis bound,'' cf.~Fig.~2.
Second, the cosmic sea of $\nu_e$'s is
predicted to be very different:  a nonthermal spectrum with higher
energy per neutrino.

Finally, we turn to model building and verification.
Within the framework of standard electroweak interactions
the decay mode $3\nu_e$ is highly forbidden, as it involves
flavor-changing neutral currents, and of course
the $\nu_e\phi$ mode does not exist.  In simple extensions of
the standard model both modes can arise.  For example, in models with
family symmetry $\phi$ is a massless Nambu-Goldstone boson
associated with the breakdown of a family symmetry (at energy scale $f$)
and $\tau_\nu \sim 8\pi f^2/m_\nu^3
\sim 10\sec (f/10^9\GeV )^2(m_\nu /10\MeV )^{-3}$.
In these models there can be many such Nambu-Goldstone
bosons which contribute additional massless degress of freedom \cite{Wilczek}.
The $3\nu_e$ decay mode can be mediated by a heavy boson with mass $M$; the
lifetime is then
$\tau_\nu \sim 0.1 {\rm sec} (m_\nu/10 {\rm MeV})^{-5}
(M/M_W)^4\sin^{-2}\theta$. We note though that in many models
this boson would
also couple to the tau lepton and violate the stringent bounds
on $\tau\rightarrow 3e$.

The best laboratory upper limits to the tau-neutrino mass are:
$31\MeV$ by the ARGUS Collaboration and $32.6\MeV$ by the
CLEO Collaboration \cite{labmass}.  They are based upon end-point
studies of the five-pion decay mode of the tau lepton.
It is possible that the mass sensitivity could be extended
to the $5\MeV - 10\MeV$ range by looking at tau decay
modes with multiple Kaons in the final state \cite{barry};
moreover, experiments done at B-factories may well be able to reach
mass sensitivities of $5\MeV - 10\MeV$ \cite{barry}.

To summarize, a tau neutrino of mass $1\MeV -10\MeV$ and
lifetime $10\sec -100\sec$ whose decay products include
electron neutrinos can significantly increase the energy
density in relativistic particles without upsetting the
successful predictions of primordial nucleosynthesis.
Alternatively, a tau neutrino in the same mass range,
but with lifetime $0.1\sec - 1\sec$, allows
many more massless degrees of freedom leading to the
same end.  The higher level of energy density in
relativistic particles modifies the CDM transfer function
fixing the nagging problem that CDM has with
excessive power on small scales.  The level of CBR anisotropy
predicted in the $\tau$CDM model is slightly larger than CDM on
the degree scale.  Most importantly, perhaps, prospects
for testing this hypothesis at $e^\pm$ colliders appear promising.

\bigskip\bigskip\bigskip
We thank L.~Hall and S.~Davidson for valuable comments about
model building.  This work was supported in part by the DOE
(at Chicago and Fermilab) and by the NASA
through grant NAGW-2381 (at Fermilab).
GG is supported in part by an NSF predoctoral fellowship.

\vfill\eject
\section*{Figure Captions}
\bigskip

\noindent{\bf Figure 1:}  The power spectra for CDM models
normalized to COBE: standard CDM ($h=0.5$); mixed dark matter
($h=0.5$); and $\tau$CDM ($h=0.45$ and $g_*=7.5$).  The
data points are from the IRAS 1.2Jy survey \cite{fisher}.
Wavenumber
and wavelength are related by $k \equiv 2\pi/\lambda$.

\medskip
\noindent{\bf Figure 2:}  Contours of the additional
number of massless species that can be tolerated without
violating the constraints to the light-element abundances
(Dirac tau neutrino and $3\nu_e$ decay mode).
[Note, because our calculations do not include inverse
tau-neutrino decays, our results {\it may} be unreliable in the lower
left-hand corner, $\tau \la (\MeV /m_\nu )^2\sec$; see
Ref.~\cite{dgt}.]

\medskip
\noindent{\bf Figure 3:}  The equivalent number of massless
neutrino species $\Delta N_\nu$ produced by the decay of a massive
tau neutrino and the excluded regions of the mass-lifetime
plane for a massive Dirac tau neutrino and the $\nu_e\phi$
and $3\nu_e$ modes.  The solid contours correspond to
$\Delta N_\nu = 4,6,8,10$ from left to right.
The excluded regions are to the right of
the broken lines (heavy for $Y_P\le 0.24$; light for $Y_P\le 0.25$).

\medskip
\noindent{\bf Figure 4:}  The COBE-normalized
angular power spectrum for standard CDM (with $h=0.5$)
and $\tau$CDM (with $h=.45$ and $\Delta N_\nu = 10$). For both,
$\Omega_B h^2=0.0125$.

\end{document}